# High-voltage Pulse Generation Based on Relaxed Self-Excited Oscillation Using Electrostatic Induction in External Capacitors

Taku SAIKI

(3-3-35 Yamate, Suita, Osaka 564-8680, JAPAN)
Department of Electrical and Electronic Engineering, Faculty, Faculty of Engineering Science, Kansai University, Japan

E-mail: tsaiki@kansai-u.ac.jp

**Keywords:** High voltage, repetitive pulses, EHD power generation, Flame jet

**Abstract**

A simple method of producing high voltage pulses with short rising times based on electrostatic induction in external capacitor used for pulse power applications is proposed. The circuit has a simple structure and contains a minimum number of parts, which makes the instrument small and light weight. In fact, the generation of sawtoothed high voltage pulses with short rising times and low repetitive rates of a few 100 Hz was successfully conducted in experiments. Theoretical analysis was simultaneously undertaken. The numerically calculated results for generating high voltage pulses were goodly consistent with the experimental ones. Moreover, it has been confirmed that amplification of the output voltage by electro-hydrodynamics (EHD) electricity generation using a jet flame resulted in lower electricity consumption and high repetition rates.

1.Introduction

Pulse power sources, which can produce high temperature or high-density extreme conditions by releasing electric or magnetic energy stored in coils or capacitors within a short time, have been introduced to various fields such as those in laser[1], fusion research[2], the production of plasma, shockwaves in water, water treatment, and exhaust gas treatments [3-13].

Various pulse power sources have now been developed due to the expanding fields of application[3-19]. The required conditions for pulse power sources differ according to applications. The required parameters for the pulse power sources of individual applications are high voltage, high current, voltage or current pulses with short standing times, large averaged output power, and high repetitive rates.

The generation of pulses using discharge switches typified by a conventional gap switch has been continuously researched [6,19]. Also, various high voltage pulse sources using insulated-gate bipolar transistor (IGBT) semiconductors with high voltage resistivity have been used as alternative switches [15-16]. The Marx bank circuit is well known as an instrument that produces voltage pulses with low repetitive rates [6]. These instruments or electrical circuits have several advantages. However, their three main problems are: 1) a complex structure, 2) high cost, and 3) excessive weight, which is very hard to reduce.

This paper proposes a new method of easily producing sawtoothed high voltage pulses with short rising times and low repetitive rates based on electrostatic induction in the external capacitor. The proposed circuit has a simple structure and no moving parts. Thus, the instrument can be small, light weight, and have a long lifetime. In addition, an experiment on high voltage pulses that were amplified due to EHD power generation using a jet flame was also carried out to find whether the electric power of the power source was reduced.

## 2. Principle for generation of high voltage pulses

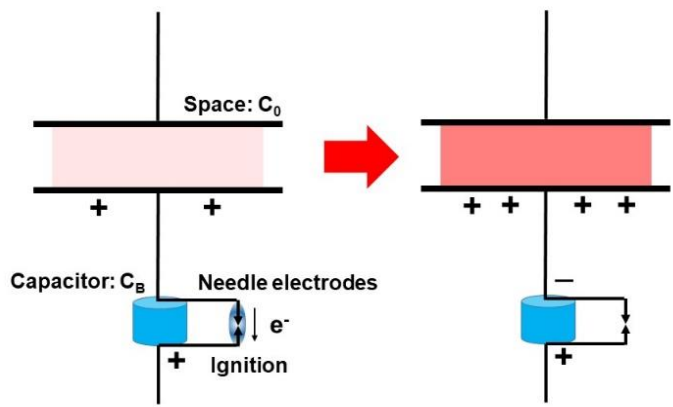

Fig. 1. Principle underlying generation of high voltage in space sandwiched by electrodes.

The principle underlying the generation of high voltage pulses by charging and discharging external capacitors is outlined below. The space between two electrodes can be seen as a capacitor with a capacitance of $C_0$ and an external capacitor with a capacitance of $C_B$, as shown in Fig. 1. Here, $C_B$ is much larger than $C_0$. We have called the external capacitor a boosted capacitor below. Here, the space between the electrodes is assumed to be filled with gas such as air or low pressure gas. A boosted capacitor is connected in series with the capacitor that consists of the space and electrodes. The boosted capacitor has needle-needle electrodes. When a high voltage is added to this circuit once, the boosted capacitor is charged to a high voltage due to electrostatic induction. The voltage of the boosted capacitor reaches the threshold voltage of the needle-needle electrodes for discharge, which occurs between the electrodes that shorten, and negative charge flows out of the boosted capacitor. Thus, a large current flows between electrodes within a short time. If the discharge has negligible energy loss, all the stored electrical energy in the boosted capacitor migrates to the capacitor, which consists of space and two electrodes. A very high voltage is generated in the capacitor, which consists of space and the two electrodes. This is because the series connected capacitors are recognized as a single capacitor with variable capacitance. The boosted capacitor is ignored by the shortening of the needle-needle electrodes when discharge occurs.

The stored electrical energy of the capacitor that consists of space and the two electrodes is explained next.

$$E_0 = \frac{1}{2} \cdot C_0 \cdot V_0^2 \quad . \tag{1}$$

Here, $V_0$ is the voltage that is added to the capacitor that consists of space and the two electrodes.

Also, the equation for the stored electrical energy of the boosted capacitor is:

$$E_B = \frac{1}{2} \cdot C_B \cdot V_B^2 \quad . \tag{2}$$

Here, $V_B$ is the voltage to the boosted capacitor when discharge occurs. After discharge has occurred, all the stored electrical energy in the boosted capacitor migrates to the capacitor that consists of space and the two electrodes.

The new electrical energy of the capacitor that consists of space and the two electrodes is next changed as:

$$E_0' = E_0 + E_B. \tag{3}$$

Thus, the equation for $V_0'$ after discharge is:

$$V_0' = \sqrt{\frac{2 \cdot E_0'}{C_0}}. \tag{4}$$

The stored charge in the boosted capacitor by the move of discharge to the capacitor consists of space and the two electrodes. After that, the moved charges slowly disappear. The equation for the increment of the stored charge at the positive electrode is:

$$\Delta Q_0 = Q'_0 - Q_0 = C_0 \cdot V'_0 - C_0 \cdot V_0 = C_0 \cdot \left\{ \sqrt{\left(\frac{C_B}{C_0}\right) \cdot V_B^2 + V_0^2} - V_0 \right\}. \quad (5)$$

Here, $Q_0 > 0$. This $\Delta Q_0$ is newly stored in the capacitor that consists of space and the two electrodes for each discharge. $\Delta Q_0(t_0)$ is the supplied charge at discharge, and $Q_0$ is calculated by:

$$\frac{dQ_0(t)}{dt} = +\Delta Q_0 \cdot \delta(t_0) - I(t), \quad (6)$$

where $t_0$ is when discharge occurs. The equation for the current that flows out of the capacitor is:

$$I(t) = \frac{V(t)}{R_p} = C_0 \cdot \frac{dV(t)}{dt}, \quad (7)$$

where $R_p$ is the internal resistance of the capacitor. The equations from (1) to (8) were converted to differential equations, and the charge, voltage, and currents of the capacitor were calculated.

The equation for the repetition time is:

$$\tau \propto \frac{C_B \cdot V_B \cdot R_S}{V_{OCW}}. \quad (8)$$

Here, $R_s$ is the charging resistance, and $V_{OCW}$ is the output voltage of the rectification circuit. The time depends on $C_B$, $V_B$, $R_s$, and $V_{ocw}$.

## 3. Method for experiment

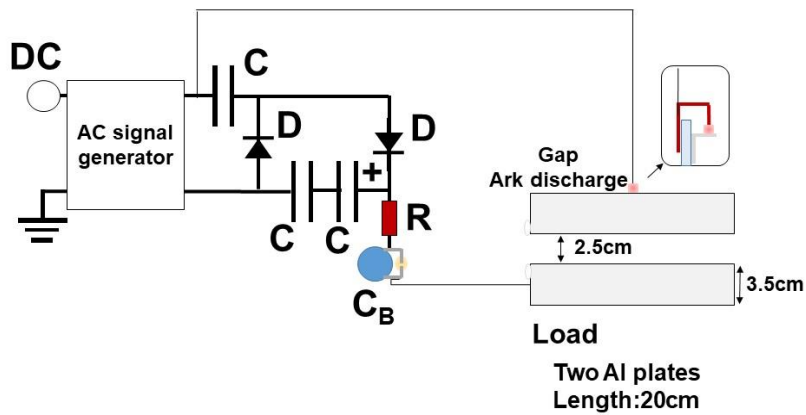

Fig. 2. Experimental setup for generating high voltage pulses.

DC: DC voltage source, $C_B$: boothed capacitor, D: high voltage withstanding diode
C: capacitor for rectification, and R: resistance for measuring current.

The experimental setup for generating high voltage pulses is outlined in Fig. 2. Two capacitors (capacitance of 100 pF (TDK, Japan), withstanding voltage of 3 kV, distance of 4 mm between needle electrodes, capacitance of 1000 pF (Murata Manufacturing, Japan), withstanding voltage of 16 kV, and a distance of 2 mm between needle electrodes) were used in the experiment as the boosted capacitors. Two diodes (Fuji Denki Systems, Japan, with a withstanding voltage of 12 kV) connected in series were used. A DC power source (PR36-3A, Texio, Japan) was also used. A high voltage generator (a harmonic power source using an ignition coil with a frequency of 20 kHz and a maximum output voltage of 5 kV) was connected to the DC power source. The high voltage generator was connected to a single-stage Cockcroft-Walton (CW) circuit (a double voltage rectifier circuit), as outlined in Fig. 1. The output was chosen on the plus electrode side of the double voltage rectifier circuit. High voltage enamel resistance with a length of 4.8 cm and resistance of 18 kohm was connected to the plus output side of the rectifier circuit to monitor the currents.

The load consisted of two aluminum plates (the capacitor consisted of space and the two electrodes. The thickness of both Al plates was 0.5 mm was used, as indicated on the right of Fig. 2. The two aluminum plates were placed vertically. Both Al plates were 3.5× 20 cm. Their edges were rounded with a diameter of 3 mm. The top of the Al plate had a needle electrode with a 1 mm gap to the bottom Al plate. Here, an arc discharge at the needle electrode by increasing the output voltage of the rectifier circuit to enhance the output voltage was needed to generate.

Here, the circuit point, at which the output of the high voltage power generator and the input of the rectifier circuit were connected, as shown in Fig. 2, was connected to the load as an output. The bottom Al plate was connected to the positive output of the rectifier circuit via the boosted capacitor.

A high voltage probe (HV-60, Sanwa, Japan), an oscilloscope (TDS-2012, Textronics, Japan), and a testa (CD731a, Sanwa, Japan) were used for measuring the voltage waveform and the average voltage was added to the load, current. The waveform of the output voltage was pulse when the load was not connected because AC and DC voltage were mixed. When the load and the boosted capacitor were connected to the rectifier circuit, arc discharge occurred at the needle gap and the load current increased, which resulted in degrading the output voltage of the DC power source and the occurrence of relaxed self-excited oscillation.

## 4．Results and Discussion

An experiment to generate high voltage pulses was conducted by using the instrument outlined in Fig. 2. The results are provided in Fig. 3.

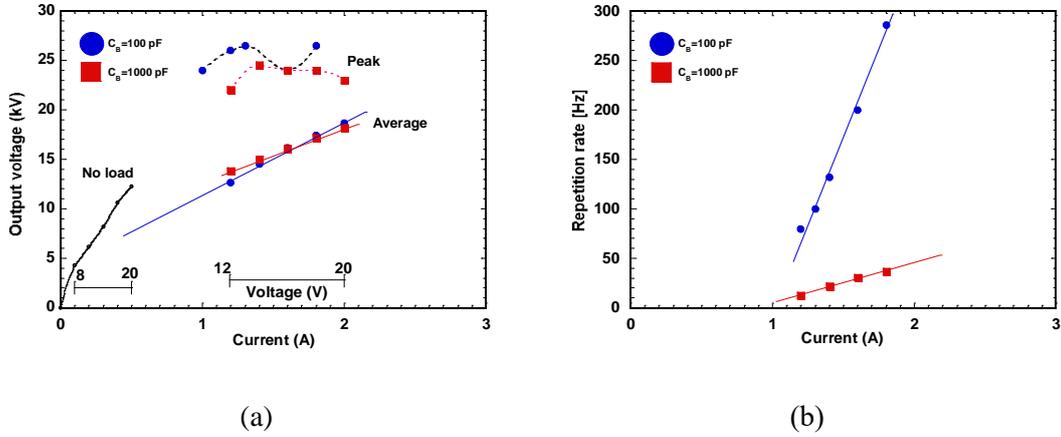

(a)                                      (b)

Fig. 3. Properties of added voltage and current applied to load.    (a) Average and peak voltage applied to load and (b) repetition rate.

The measured voltage added to the load and current are given in Fig. 3(a). The output voltage of the rectifier circuit was proportionally changed to the output voltage or current of the DC power source when no arc discharge was generated at the needle gap. A maximum average output voltage of 12.5 kW was obtained for an output voltage of 20 V and an output current of 0.5 A. The load current increased and the output voltage of DC power supply degraded when arc discharge at the needle gap was generated and $C_B$ =100 pF was used. Average output voltages of 12 kV and 19 kV were obtained for the output currents of 1.2 and 2.0 A, as shown in Fig. 3(a). An output peak voltage of around 26 kV was obtained.

The load current increased, and the output voltage of the DC power supply degraded when arc discharge was generated at the needle gap and $C_B$ =1000 pF was used. Average output voltages of 13 and 18 kV were obtained for the output currents of 1.2 and 2.0 A, as shown in Fig. 3(a). An output peak voltage of around 25 kV was obtained. The repetition rates obtained from the oscilloscope are plotted in Figs. 3(a and b)

The repetition rate changed from 80 and to 290 Hz for the output currents of 1.2 and 2.0 A when $C_B$ =100 pF was used, as shown in Fig. 3(b), whereas the repetition rate changed from 10 and to 40 Hz for the output currents of 1.2 and 2.0 A when $C_B$ =1000 pF was used, as shown in Fig. 3(b), due to the increasing charge time of the boosted capacitor.

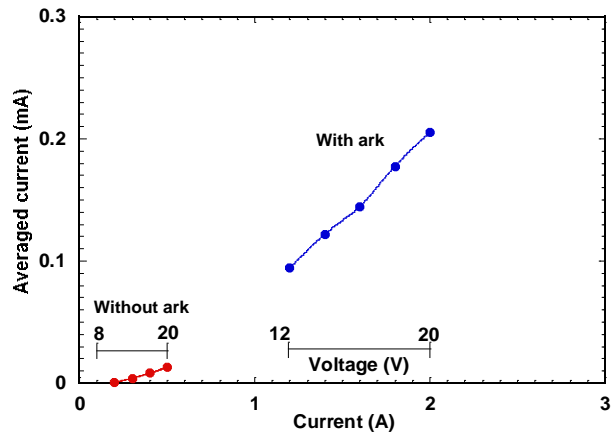

Fig. 4. Measured average output current.

Fig. 4. Measured average output current for rectification circuit. Source current was changed from 1.2 to 2.0 A for $C_B$ =100 pF, and average output current for rectification circuit was changed from 0.1 to 0.21 mA. Average output current increased by 20 times compared to when there was no load.

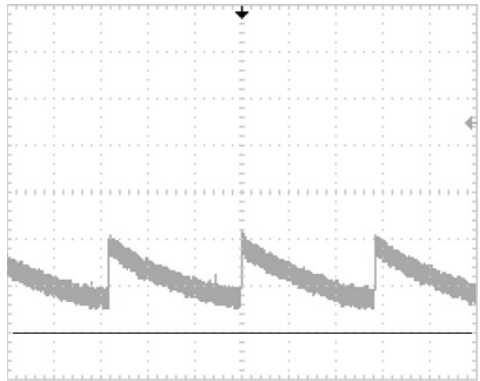

Fig. 5. (a) Waveforms of applied voltage to load. Without arc discharge, current: 0.4 A, repetition rate: 7.1 Hz, $C_B$ =100 pF, 5 kV/div., and 50 ms/div.

The waveforms of applied voltage to the load obtained by oscilloscope at $C_B$ = 100 pF and without arc discharge are given in Fig. 5. (a), where the 20 kHz AC component appeared in the waveform. Sawtoothed high voltage pulses applied to the load with the repetition rate of 7.1Hz were observed. The voltage increments applied to the load within a short time were 6 kV when discharge occurred at the needle electrodes of the boosted capacitor. After that, the voltages applied to the load slowly decreased due to outflowing stored charges.

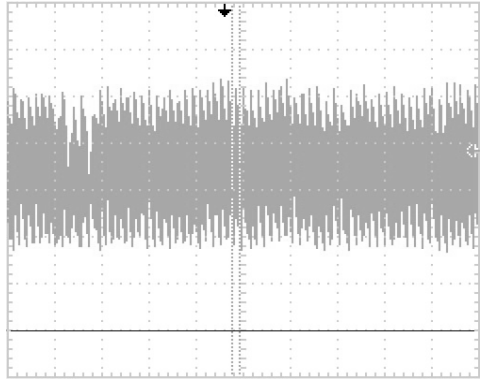

Fig. 5. (b) Waveforms of applied voltage to load. Current: 1.8 A, repetition rate: 290 Hz, 5 kV/div, and 50 ms/div.

The waveforms of applied voltage to the load obtained by oscilloscope at $C_B$ = 100 pF and with arc discharge are provided in Fig. 5. (b). Also, sawtoothed high voltage pulses applied to the load with a repetition rate of 290 Hz were observed, where the output current increased to 1.8 A and blocking oscillation was observed after arc discharge was generated. The voltage increment applied to the load within a short time was 4 kV when discharge occurred at the needle electrodes of the boosted capacitor.

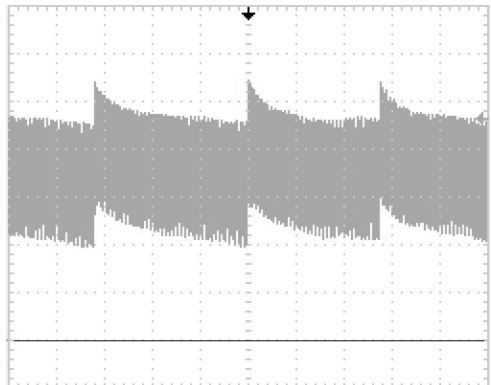

Fig. 5. (c) Waveforms of applied voltage to load. Current: 2.0 A, repetition rate: 33 Hz, 5 kV/div, and 10 ms/div.

The waveforms of the applied voltage to the load obtained by oscilloscope at $C_B$ = 1000 pF and with arc discharge are given in Fig. 5. (c). Also, sawtoothed high voltage pulses applied to the load with a repetition rate of 33 Hz were observed, where the output current increased to 2.0 A and blocking oscillation was also observed after arc discharge was generated. The voltage increment applied to the load within a short time was 5 kV when discharge occurred at the needle electrodes of the boosted capacitor. Variations occurred in the voltage applied to the load by changing the space gap of the load. However, there was less change in the average or peak voltage applied to the load and the repetitive rate was observed in an experiment.

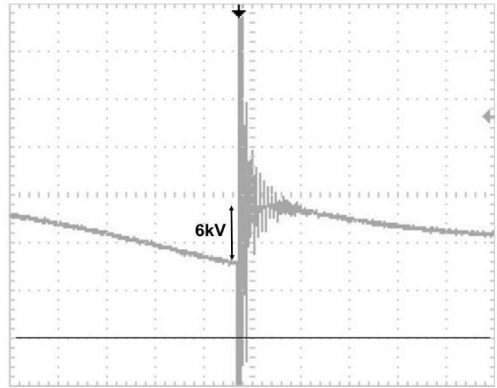

Fig. 5. (d) Waveform of applied voltage to load: $C_B = 100$ pF, 5 kV/div, and 1 μs /div.

The waveform of the standing voltage applied to the load at $C_B =100$ pF is shown in Fig. 5. (d). It was found that the standing time was a few tens of nanoseconds. The waveform then indicated an oscillation due to the inductance component. Here, the increment of voltage was 6 kV. The waveform of the standing voltage applied to the load observed by using a spherical electrode gap at $C_B =1000$ pF is given in Fig. 5. (e). We found that the standing time was the same, and the waveform indicated an oscillation due to the inductance component. Here, the increment of voltage was 15 kV.

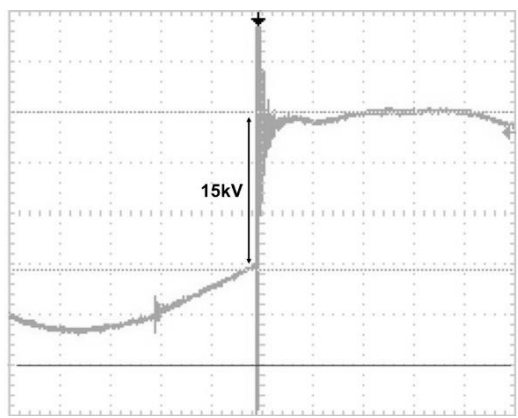

Fig. 5 (e). Waveform of applied voltage to load: as an example, 1000 pF, sphere gap electrode: 2 mm gap (discharge voltage: 6 kV), 5 kV/div, and 2.5 μs/div.

The waveform of current at $C_B =100$ pF is indicated in Fig. 5 (f), as an example. This was because the waveform of the output voltage was a pulse, and the current flowed asymmetrically. A larger current flowed when the voltage was high. A pulse current simultaneously flowed in the load at the discharge of the needle electrode gap. The current was evaluated to be 22 mA, which was determined from the voltage of 400 V at the waveform, as shown in Fig. 5 (e). The duration of the discharge pulse was evaluated to be 3.3 µs.

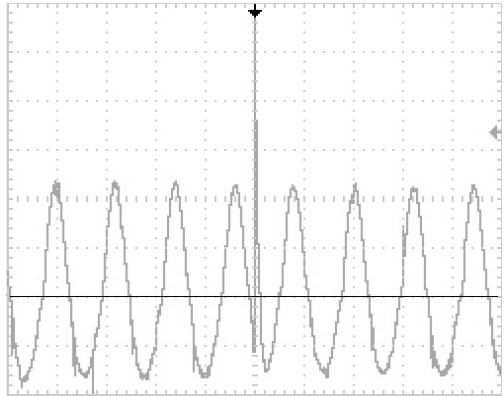

Fig. 5 (f). Waveform of current: as an example, output current of DC power source: 1.2 A, 5 kV/div, and 25 µs/div.

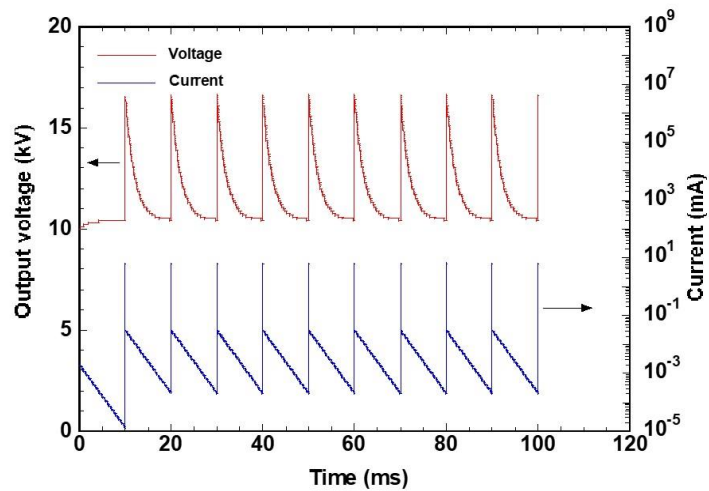

Fig. 6. Numerically calculated waveform of voltage applied to load, as an example.

The numerically calculated waveform of the voltage applied to the load, as an example, is shown in Fig. 6. Equations (5) to (7) were used for calculating the voltage applied to the load and current and charge, where the repetition rate is 100 Hz for the calculation parameter. The $V_B$ is 4 kV, $C_0$ is 10 pF, $C_B$ is 100 pF, $V_{ini}$ is 10kV, and $R_p$ is 200 Mohm. The time mesh was set to 10 μs. The calculated results in Fig. 6 are the sawtooth waveforms and are consistent with the results in Fig. 5. It was found that the output currents that flowed out of the load were 6.5 mA at peak and there was a weak current of below 0.1 mA at the charging boosted capacitor.

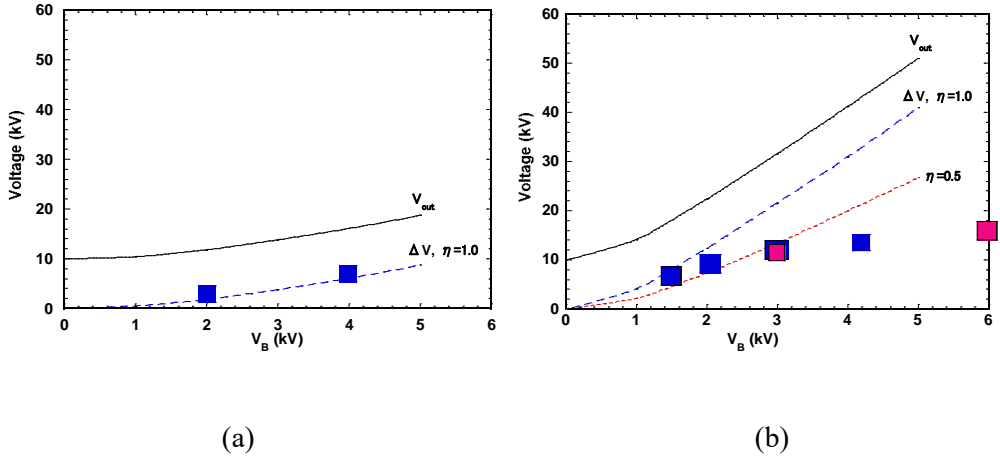

(a)          (b)

Fig. 7. Voltage applied to load and raised voltage of pulse during discharge for $V_B$. (a) $C_B$ =100 pF and (b) $C_B$ =1000 pF. Red squares plot experimental results when spherical electrode gap was used, and blue squares plot experimental results when needle electrode gap was used.

The calculated and measured raised voltage of the pulse during discharge for $V_B$ is shown in Fig. 7. Equations (1) to (4) were used for calculating the voltage applied to the load. The solid line plots the calculated voltage applied to the load, and the dashed line indicates the calculated raised voltage of the pulse during discharge. Also, $\eta$ in Fig. 7(b) is the energy transfer efficiency from the boosted capacitor to the load. The experimental results are goodly consistent with the numerically calculated ones at $C_B$ =100pF, as shown in Fig.7(a). However, the experimental results are not consistent with the numerically calculated ones at $C_B$ =1000 pF, as shown in Fig. 7(b). The measured $V_B$ reduced from the calculated ones at $\eta$=1.0. Because the experimental results between using the needle electrode gap and using the spherical electrode gap were almost the same, we expected this to be due to a reduction in capacitance in the ceramic capacitor at high voltage because of material properties. The measured $V_B$ was close to the calculated one at $\eta$=0.5. It was found that the stored electrical energy of 0.8 mJ completely migrated to the load at $C_B$ =100 pF.

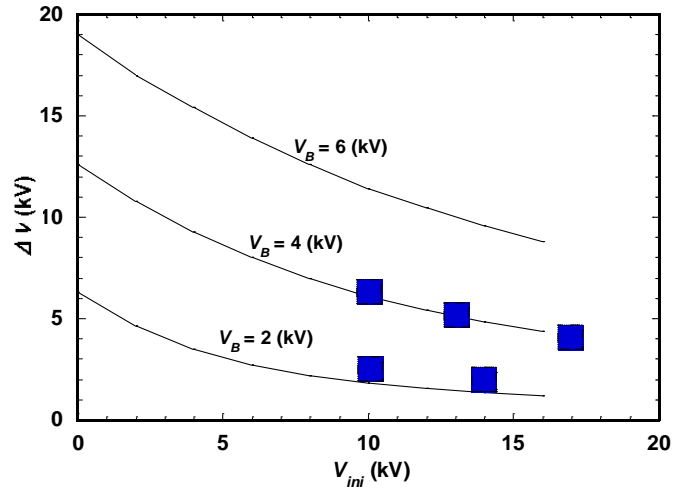

Fig. 8. Raised voltage of pulse during discharge for $C_B$. $C_B$ =100 pF and $C_0$ =10pF. Solid line plots calculated applied voltage to load. Blue squares plot experimental results when using needle electrode gap.

The raised voltage of the pulse during discharge for $C_B$ are plotted in Fig. 8. Equations (1) to (4) were also used for calculating the voltage applied to the load. It was recognized that the charges in the load depended on the initial voltage, $V_{ini}$, that was applied to the load. If $C_0$=10 pF, $C_B$=100 pF, $C_B$=4 kV, and $V_{ini}$ is 0 kV, the boosted capacitor has 32% of the charge stored in the load. If $V_{ini}$ is 10 kV, only 15% of the charge is stored in the load. The DC component of the output voltage for the rectification circuit must remain low with increasing repetition rate.

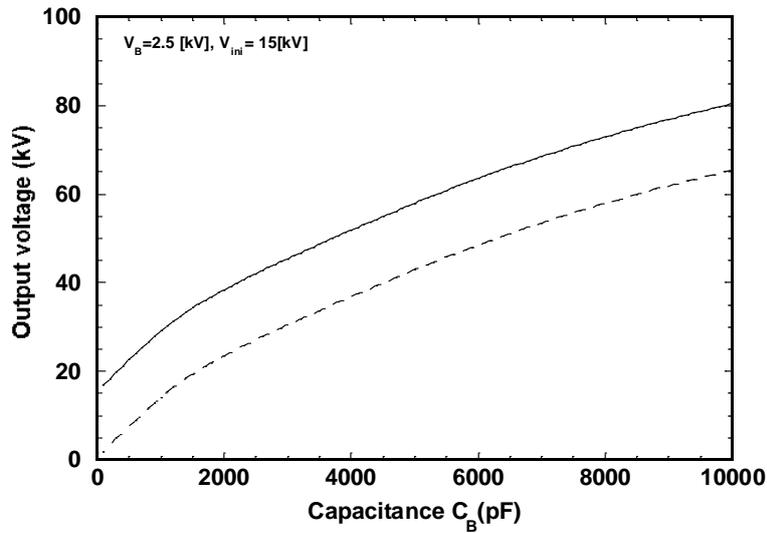

Fig. 9. Applied voltage to load and raised voltage of pulse during discharge as function of $C_B$. Solid line plots applied voltage to load. Dashed line plots raised voltage of pulse during discharge.

The numerically calculated results for the raised voltage of the pulse during discharge are plotted in Fig. 9, where $C_B$ =2.5 kV and $V_{ini}$ =15 kV. When capacitance increased, the raised voltage of the pulse became saturated. It has been recognized that we needed to use a capacitor with lower capacitance and high withstand voltage instead of using a capacitor with high capacitance.

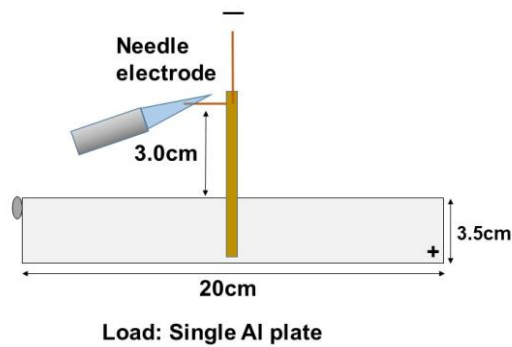

Fig. 10. Load using flame jet EHD power generation.

Flame jet EHD power generation [20] was used to enhance the output voltage and reduce consumption power. The structure of the load was modified to the single Al electrode in Fig. 10 to adapt the experiment using flame jet EHD power generation. A negative needle electrode was used for the load, and the flame coming out from a butane burner (flame length: 5 cm) was closed, as shown in Fig. 10. The distance between the needle electrode and Al electrode was set to 3.0 cm. When $C_B$ =100 pF, the repetition rate was remarkably increased from 17 Hz to 250 Hz and a corona discharge from the Al plate electrode was observed. The generated voltage between the flame and the Al plate electrode of the load was estimated to be 40 kV, which was estimated from the corona discharge length of 1.5 cm. A maximum amplified voltage of 40 kV was also prospected from the prevous experiment [20].

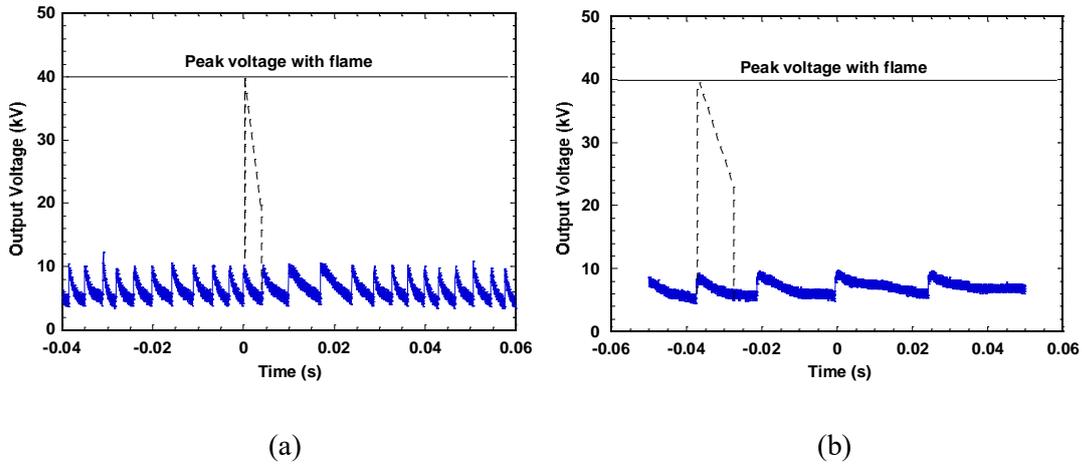

Fig. 11. Applied voltage to load. (a) $C_B$ =100 pF, current of power source: 0.5 A, and repetition rate: 130 Hz, (b) $C_B$ =1000 pF, current of power source: 0.5 A, and repetition rate: 250 Hz.

The measured voltage applied between electrodes is plotted in Fig. 11(a). The output current of the DC power source was 0.5 A at $C_B$ =100 pF, and the repetition rate increased to 250 Hz. The average voltage between electrodes was 7 kV. The output current of the DC power source was 0.5 A at $C_B$ =1000 pF, and the repetition rate increased to 130 Hz. The average voltage between electrodes was 6 kV.

The repetition rate for the output current of the DC power source is plotted in Fig. 12(a). The repetition rate at $C_B$ =100 pF changed to a few Hz and 20 Hz for each output current of 0.3 and 0.5A when the flame jet was not used. The repetition rate increased to 200 and 250 Hz for each output current of 0.3 and 0.5A when the flame jet was used. The repetition rate at $C_B$ =1000 pF changed to 1.4 and 2.6 Hz for each output current of 0.3 and 0.5A when the flame jet was used. The repetition rate also increased to 16 and 130 Hz for each output current of 0.3 and 0.5A when the flame jet was used.

The measured average output current for the rectification circuit is shown in Figs. 12(b) and (c). The output current of the DC power source at $C_B$ =100 pF is 0.5A, and the average output current increased to 0.32 mA. The output current of the DC power source is 0.5A at $C_B$ =1000 pF, and the average output current increased to 0.28 mA.

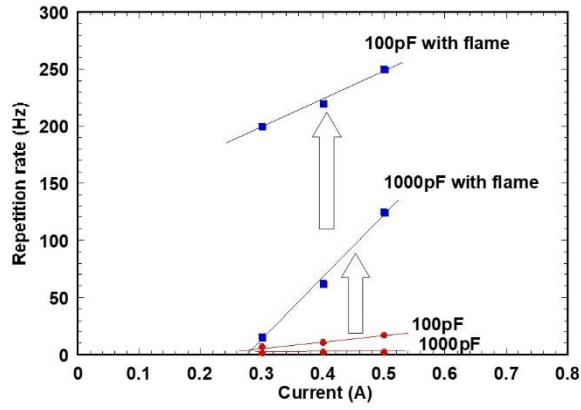

(a)

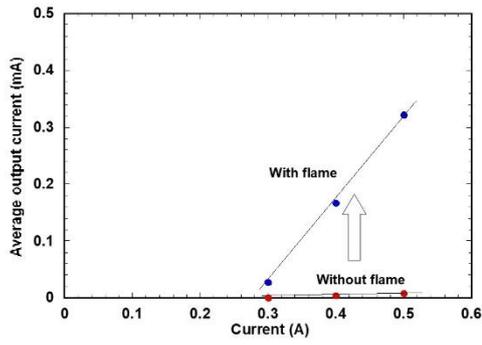

(b)

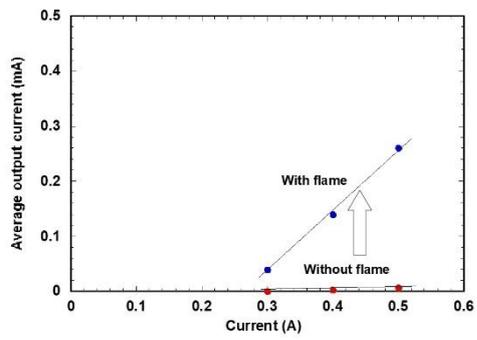

(c)

Fig. 12. Properties of circuit. (a) repetition rate, (b) average current: $C_B$ =100 pF , and (c) average current: $C_B$ =1000 pF.

The results obtained from the experiments are summarized in Table 1.

Table 1. Comparisons of high voltage pulses being generated using relaxed self-excited oscillation and that using flame jet EHD power generation.

| | Output power of supply (W) | Total output power applied to load (W) | Power from boosted capacitor to load (W) |
|---|---|---|---|
| Using only relaxed self-excited oscillation | 40 | 4 | 0.2 |
| Using flame jet EHD power generation | 10 | 2.1 (only rectification circuit) <br><br> 9 (with flame) | 5 (0.3 only rectification circuit) |

This method easily produced high voltage pulses from a DC power source. The limit of the repetition rate of the voltage pulse should be 1 kHz due to the use of the discharge gap. Part of the output electrical power for the rectification circuit migrated to the boosted capacitor, and the stored electrical energy was injected to the load with a different timing.

The average voltage added to the load was 7 kV when flame jet EHD power generation was used and the average current was 0.32 mA, where the consumption power was evaluated to be 2.1 W. The peak voltage of the load due to voltage amplification of the flame jet EHD increased to around 40 kV. This total consumption power at load was evaluated to be 9 W. When the $C_B$ was 100 pF, the stored electrical energy was 0.8mJ, the repetition rate was 250 Hz, and the transfer electrical energy from the boosted capacitor to the load was evaluated to be 0.2 W. When the $C_B$ was 1000 pF, the stored electrical energy was 2.4 mJ, the repetition rate was 130 Hz, and the transfer electrical energy from the boosted capacitor to the load was evaluated to be 0.3 W. It was found that the transfer electrical power in both cases was almost the same. The voltage added to the load was multiplied by four times due to the voltage amplification of flame jet EHD, and the transfer electrical power to the load was amplified by 16 times the original power. Thus, the transferred electrical power was finally increased to 5 W, which was the effect of the voltage amplification of flame jet EHD. The multiplication was remarkably significant.

The maximum power from the DC power supply was 10 W. However, we could not obtain adequate capabilities from the rectification circuit because the load was very small. If we had used load with high levels of current, the energy efficiency of the rectification circuit should have been improved. The three main contributions to improvements were: 1) using fewer variations in capacitance for the boosted capacitor, 2) using a flat capacitor with reduced weight, and 3) using a small flame source. Thus, using these would reduce the weight of the instrument.

4. Conclusion

A simple method of producing high voltage pulses based on electrostatic induction without moving parts was introduced. Sawtoothed voltage pulses with short standing times and repetitions of a few 100 Hz were actually produced in this experiment. The experimental results for generating the voltage pulses were goodly consistent with the numerically calculated ones. We demonstrated that high voltage pulses could be produced by a combination of flame jet induced EHD power generation and a boosted capacitor with low consumption of electricity in the load. The power consumption was one order lower than that using normal relaxed self-excited oscillation, and the transferred electrical energy from the boosted capacitor to the load was multiplied to 5 W.

This research did not receive any specific grant from funding agencies in the public, commercial, or not-for-profit sectors.

**References**
[1] C. K. Patel, "Continuous-Wave Laser Action on Vibrational-Rotational Transitions of CO2". Physical Review, **136** (5A) (1964) A1187–A1193.
[2] M. K. Matzen, "Z pinches as intense x-ray sources for high-energy density physics applications", Physics of Plasmas, **4**（1997）1519.
[3] H. Akiyama, S. Sakai, T. Sakugaw, and T. Namihira, "Environmental applications of repetitive pulsed power", IEEE Transactions on Dielectrics and Electrical Insulation, **14** (2007) 825.
[4] B. Sun, M. Sato, "Use of a pulsed high-voltage discharge for removal of organic compounds in aqueous solution", J. Phys. D: Appl. Phys., **32** (1999), pp. 1908-1915
[5] J. S. Clements, M. Sato, R.H. Davis, "Preliminary investigation of prebreakdown phenomena and chemical reactions using a pulsed high-voltage discharge in water", IEEE Trans. Ind. Appl., 23 (1987), pp. 224-235
[6] I.V.Lisitsyn, H.Nomiyama, S.Katsuki and H.Akiyama, "Thermal Processes in a Streamer Discharge in Water, "IEEE Trans. Dielectr. Electr. Insulat., **6**(3), (1999) 351.
[7] M. Rezal, Dahaman Ishak, M. Sabri, "High voltage magnetic pulse generation using capacitor discharge technique", Alexandria Engineering Journal, **53** (2014), pp803–808.
[8] T. Shao, G.S. Sun, P. Yan, S.C. Zhang, Breakdown phenomena in nitrogen due to repetitive nanosecond-pulse, IEEE Trans. Dielectr. Electr. Insul., **14** (2007) pp.813–819.

[9] E.L. Neau, Environmental and industrial applications of pulsed power systems, IEEE Trans. Plasma Sci., **22** (1994) pp.2–3.

[10] F. Fukawa, N. Shimomura, T. Yano, Application of nanosecond pulse power to ozone production by streamer corona, IEEE Trans. Plasma Sci., **36** (2008)pp.2592–2597.

[11] Fazhi Yan, Baiquan Lin, Chuanjie Zhu, Yan Zhou, Xun Liu, Chang Guo, Quanle Zou, "Experimental investigation on anthracite coal fragmentation by high-voltage electrical pulses in the air condition: Effect of breakdown voltage", Fuel, **183** (2016) pp.583–592.

[12] S. Katsuki, H. Akiyama, A. Abou-Ghazala and K.H. Schoenbach, "Parallel Streamer Discharges Between Wire and Plane Electrodis in Water", IEEE Trans. Dielectr. Electr. Inslat. 9, (2002) 498.

[13] M. Sato, T. Tokutake, T. Ohshima and A.T. Sugiarto, ," Aqueous Phenol Decomposition by Pulsed Discharges on the Water Surface", IEEE Trans. Industry Applications 44, (2008) 1397.

[14] H. Li, A. Lukanin, A. Tskhe, S. Sosnovskiy, "Multifunctional generator of high-voltage microsecond pulses", Journal of Electrostatics, **90** (2017) pp74–78.

[15] T. Shao, et al., Excitation of atmospheric pressure uniform dielectric barrier discharge using repetitive unipolar nanosecond-pulse generator, IEEE Trans. Dielectr. Electr. Insul., **16** (2010) pp1830–1835.

[16] T. Shao, D.D. Zhang, Y. Yu, A compact repetitive unipolar nanosecond-pulse generator for dielectric barrier discharge application, IEEE Trans. Plasma Sci., **38** (2010)pp.1651–1652.

[17] K. Yan, E. J. M. van Heesch, A. J. M. Pemen, P. A. H. J. Huijbrechts, F. M. van Gompel, H. van Leuken, and Zdenek Matyáˇs, "A High-Voltage Pulse Generator for Corona Plasma Generation", IEEE Transcations on Industry Applications, **38**(3) (2002) pp.866-872.

[18] D. Wang, S. Okada, T. Matsumoto, T. Namihira and H. Akiyama, "Energy transfer efficiency of nano-seconds pulsed power generator for nonthermal plasma processing technique", IEEE Plasma Sci., **38**, 2746 (2010).

[19] J.-D. Moon, "A compact high-voltage pulse generator using a rotary airhole sparkgap", Journal of Electrostatics, **65** (2007) 527–534.

[20] T. Saiki, "Study on High Voltage Generation Using Flame Column and DC Power Supply", J of Electrostatics, **70** (2012) pp.400-406 .